\documentclass[gamm,a4paper,fleqn%
]{w-art}
\usepackage{times,cite,w-thm}
\usepackage{amsmath,amssymb}
\theoremstyle{plain}

\theoremstyle{definition}

\newcommand\la{\lambda}
\newcommand\al{\alpha}

\renewcommand{\matrix}[2]{\left( \!\! \begin{array}{#1} #2 \end{array} \!\! \right)}

\newcommand{\N}{{\mathbb N}}

\newcommand{\R}{{\mathbb R}}
\newcommand{\C}{{\mathbb C}}

\renewcommand\Re{{\rm Re\,}}
\renewcommand\Im{{\rm Im\,}}

\newcounter{marke}
\newcommand{\bl}{\begin{list}{\roman{marke})}{\usecounter{marke}
\topsep 0 cm \itemsep 0cm}}
\newcommand{\el}{\end{list}}

\usepackage[]{graphicx}

\begin{document}
\DOIsuffix{theDOIsuffix}
\Volume{29}
\Month{01}
\Year{2007}
\pagespan{1}{}
\Receiveddate{XXXX}
\Reviseddate{XXXX}
\Accepteddate{XXXX}
\Dateposted{XXXX}



\title[On a spectral problem in magnetohydrodynamics]{On a spectral problem in magnetohydrodynamics \\
and its relevance for the geodynamo}

\author[F. Stefani]{Frank Stefani\inst{1}%
      \footnote{E-mail:~\textsf{F.Stefani@hzdr.de}}}
\address[\inst{1}]{Helmholtz-Zentrum Dresden-Rossendorf, Bautzner 
Landstr.\ 400,
01328 Dresden, Germany}

\author[C. Tretter]{Christiane Tretter\inst{2,}%
  \footnote{Corresponding author\quad E-mail:~\textsf{tretter@math.unibe.ch}}}
  \address[\inst{2}]{Universit\"at Bern, Mathematisches Institut (MAI),
Sidlerstr.\ 5, 3012 Bern, Switzerland}

\begin{abstract}
  One of the most remarkable features of the geodynamo is the 
  irregular occurrence of magnetic field reversals. Starting with 
  the operator theoretical treatment of a non-selfadjoint 
  dynamo operator, we elaborate a dynamical picture of those
  reversals that rely on the existence of exceptional spectral 
  points. 
\end{abstract}
\maketitle                   

\section{Introduction}

Probably since more than 4 billion years, the liquid metal 
flow in the Earth's outer core has been working as
a homogeneous, hydromagnetic dynamo that 
converts gravitational and thermal energy into magnetic
field energy \cite{MERRILL}.
One of the most impressive features of the geomagnetic field
is the irregular occurrence of polarity reversals, with a 
mean reversal rate of approximately 4-5 per Myr. 

Recent numerical simulations of the coupled system of the
Navier-Stokes equation for the velocity field 
and the induction equation for the magnetic field
have been successful in
reproducing not only the dominance of the axial dipole     
but also  polarity reversals \cite{GLARO}. 
Reversals were also
observed in one \cite{BERHANU} of the recent liquid sodium 
dynamo experiments which have flourished during 
the last two decades \cite{ZAMM}.
However, neither in simulations 
nor in experiments it has been
possible to accommodate all dimensionless 
parameters of the geodynamo \cite{GUBBINS}.

A complementary approach to reversals
investigates simplified,  
lower-dimensional models, with the underlying hope 
to capture the most 
essential characteristics of the
reversal process. Perhaps the most simple reversal 
model of this kind was proposed by P\'etr\'elis et al. \cite{PETRELIS} 
in form of the ordinary differential equation
$\dot{\Theta}=a_1+a_2\sin(2 \Theta) +\Delta \zeta(t)$, 
where $\Theta$ is related to the angle between the
dipole and the quadrupole components of the field,
and $\Delta$ is the amplitude of some noise.
The dynamics of this system is governed by the
existence of two pairs of stable and unstable
fixed points, or a limit cycle for other  
parameters $a_1$ and $a_2$.
Even below the saddle-node bifurcation, 
the noise term proportional to $\Delta$ can make the 
system run from one stable fixed point, via the nearest
unstable fixed point, towards the opposite stable fixed
point, thereby realizing a reversal.

Interestingly, this typical dynamical systems picture has 
an equivalent counterpart in the spectral theory
of the dynamo operator that governs the magnetic field 
dynamics. The transitions between non-oscillatory
and oscillatory eigenmodes, which are often observed here
\cite{PARKER,DEIN,YOSHI,DUDL,SAJO}, 
are well-known in operator theory as spectral 
branch points -- ``exceptional points'' of branching type of 
non-selfadjoint operators \cite{KATO}. Such 
branch points are characterized not only
by coalescing eigenvalues but also by a coalescence 
of two or more
eigenvectors within a purely geometric eigenspace 
and the formation of a non-diagonal Jordan block structure with associated vectors 
\cite{CZECH1,SEYR}.
This is in contrast to ``diabolical points'' \cite{BERRY} which
are  points of an accidental crossing of two or 
more spectral branches with an
unchanged diagonal block structure of the operator 
and without coalescing eigenvectors  \cite{KATO,SEYR}.

In a series of papers \cite{PRE,PRL,EPSL,GAFD,LUCA,INVERSE}, 
the magnetic field dynamics in the vicinity of an exceptional point 
was analyzed in greater detail numerically. For this purpose 
a so-called mean-field dynamo 
model of $\alpha^2$ type was utilized, with a supposed spherically
symmetric helical turbulence function $\alpha$, which 
is, admittedly, far away from the reality of the
Earth's dynamo. 
Remarkably, this simple model 
exhibited quite a number of the typical reversal
features, in particular a temporal asymmetry 
with slow decay and fast recovery. 
However, since the 
eigenvalue problems underlying these time-evolutions 
are not symmetric, corresponding numerical computations
are prone to be unreliable and thus have to be interpreted with care \cite{TE05, Boegli, BST18}. 
Moreover, there are questions that can never be 
answered by numerical computations, e.g.\ are there 
only finitely many non-real eigenvalues?

In the present paper, we will summarize and update
this conceptual relation between the spectral theory
of non-selfadjoint dynamo operators and the
theory of geodynamo reversals. After a short general 
introduction into dynamo theory, we will 
streamline the operator-theoretical treatment of
spherically symmetric $\alpha^2$ dynamos, as elaborated in
\cite{UWE1,UWE2}, with focus on some strict global estimates 
for the eigenvalues of these operators, 
including an anti-dynamo theorem and 
a result on the finiteness of the number of non-real eigenvalues.
Then we will move on to the temporal magnetic field evolution, 
which takes into account
the dynamical back-reaction of the magnetic field on 
the source of its generation (i.e.\ $\alpha$). 
Here we will show 
how the various features of geodynamo reversals
trace back to the spectral theory of the underlying 
operators.  The paper will close with an outlook 
on future developments.

\section{Theoretical basis of dynamo theory}

Dynamo theory starts with
Amp\`ere's law, Faraday's law, and Ohm's law 
in electrical conductors moving with velocity ${\bf{v}}$,
\begin{align}
\nabla \times {\bf{B}}&= \mu_0 {\bf{j}}\:, \label{1} \\
\nabla \times {\bf{E}}&= -\dot{{\bf{B}}}\:, \label{2} \\
{\bf{j}}&=\sigma ({\bf{E}}+{\bf{v}} \times {\bf{B}})\:, \label{3} 
\end{align}
where $\sigma$ denotes the conductivity of the fluid and
$\mu_0$ is the permeability of the vacuum  (we suppose
all materials to be non-magnetic).
We have skipped the displacement current
in Eq.\ (\ref{1}) since the quasi-stationary approximation
is usually fulfilled in good conductors.                    
Taking the $curl$ of Eqs.\ (\ref{1}) and (\ref{3}), and
inserting into
Eq.\ (\ref{2}), one readily arrives at 
the compact {\it induction equation}
for the magnetic field,
\begin{eqnarray}
\frac{\partial {{\bf{B}}}}{\partial t}=\nabla
\times ({\bf{v}} \times {\bf{B}})
+\frac{1}{\mu_0 \sigma} \Delta {\bf{B}} \label{4} \; .
\end{eqnarray}
In deriving Eq.\ (\ref{4}) we have assumed, for simplicity,
$\sigma$ and $\mu$ to be  constant in the considered region
(generalizations can be found in \cite{GIESECKE}), and we
have exploited the fact that
the magnetic field is source-free,
\begin{eqnarray}
\nabla \cdot {\bf{B}}=0 \label{4a} \; .
\end{eqnarray}
Assuming that there are no excitations of the            
magnetic field from outside, i.e.\ the field is being 
{\it self-excited} by the dynamo process,           
the boundary condition becomes                
\begin{eqnarray}                                                  
{\bf{B}}={\bf{O}}(r^{-3}) \; \;\; \mbox{as}                         
\; \;\; r \rightarrow \infty \label{3a} \; .                       
\end{eqnarray}

Obviously, the evolution of the magnetic 
field in Eq.\ (\ref{4})
is governed by the competition between generation and
diffusion
of the field. For vanishing velocity
the magnetic field will disappear within a typical decay
time $t_d= {\mu_0 \sigma} l^2$,
with $l$ being a typical length scale of the system.
When switched on, the advection by the velocity field $\bf{v}$ 
can lead to an increase of
${\bf{B}}$ within a kinematic time $t_k=l/v$, 
where $v$ is a typical velocity scale. If this
kinematic time
becomes smaller than the diffusion time, the net effect of
the evolution can become positive, leading to a growing 
magnetic field.      
Comparing the diffusion time-scale with the kinematic 
time-scale, we get a dimensionless number that 
governs the ``fate'' of the
magnetic
field. This number is called the magnetic 
Reynolds number $R_m$,
\begin{eqnarray}
R_m=\mu_0 \sigma l v \; .
\end{eqnarray}
Depending on the specific flow pattern,                                  
the critical values of 
$R_m$ for dynamo action to set in 
are typically in the range of 10$^1$ to 10$^3$. 

An important feature of dynamo theory 
concerns the assumptions on the velocity field. If we assume the
velocity to be steady, the time dependence of ${\bf{B}}$        
in Eq.\ (\ref{4}) becomes exponential according to
${\bf{B}}({\bf{r}},t)\!=\!\exp({\lambda t}) \hat{\bf{B}}({\bf{r}})$,
and the corresponding {\it kinematic} 
dynamo problem can~be rewritten as a time-independent
eigenvalue equation. 
The eigenvalue parameter $\lambda\!=\!p\!+\!2 \pi {\rm i} f$ is complex, with 
growth rate $p$ and frequency $f$.
The onset of dynamo action occurs for 
$\Re \lambda= p \ge 0$. 
However, according to Lenz's rule,
the velocity field being the source of dynamo action 
cannot remain unaffected 
when the magnetic field
grows. The increasing magnetic field
induces an increasing Lorentz 
force $\bf{j} \times \bf{B}$ which acts back on the
velocity field.
In general, the evolution of the velocity field is 
governed by the
Navier-Stokes equation
\begin{eqnarray}
\frac{\partial {{\bf{v}}}}{\partial t}+({\bf{v}}
\cdot \nabla) {\bf{v}}=
- \frac{\nabla p}{\rho} + \frac{1}{\mu_0 \rho}
(\nabla \times {\bf{B}})
\times {\bf{B}}+\nu \Delta  {\bf{v}}+{\bf{f}}_{d} \label{7} \; ,
\end{eqnarray}
where  $\rho$ and $\nu$ denote
the density and the kinematic viscosity of the fluid, respectively,
$p$ is the pressure,
and ${\bf{f}}_{d}$ symbolizes driving forces.
The Lorentz force in Eq.\ (\ref{7}) puts Lenz's rule into action:
the magnetic field acts back
on the flow, in general weakening the source of its own generation.        
The treatment of the coupled system of Eqs.\ (\ref{4}) and
(\ref{7}),
which is called the {\it dynamically consistent} dynamo problem,                      
is much more intriguing than the treatment of the linear
induction equation alone.
  
A second characteristic feature is whether or not the 
turbulent character of practically all dynamo 
related flows is reflected by the dynamo model. 
{\it Laminar  models} are described by the unchanged                         
Eq.\ (\ref{4}) with neglected turbulence.
The self-excited magnetic field varies on
the same
length scale as the velocity field does.
{\it Mean-field dynamo models}, on the other hand, are relevant for
highly turbulent flows.
In this case the velocity and the magnetic field are
considered as superpositions of mean and fluctuating parts,
${\bf{v}}=\overline{\bf{v}}+{\bf{v}}'$ and
${\bf{B}}=\overline{\bf{B}}+{\bf{B}}'$. From Eq.\ (\ref{4}) we             
obtain the
equation
for the mean part $\overline{\bf{B}}$,
\begin{eqnarray}
\frac{\partial {\overline{\bf{B}}}}{\partial t}=\nabla
\times (\overline{\bf{v}} \times \overline{\bf{B}}+{\bf{\cal{E}}})
+\frac{1}{\mu_0 \sigma} \Delta \overline{\bf{B}} \label{8} \; .
\end{eqnarray}
Obviously, the equation for the mean-field is identical
to the equation for the original field, except for one additional      
term
\begin{eqnarray}
{\bf{\cal{E}}}=\overline{{\bf{v}}' \times {\bf{B}}'} \; ,\label{9}
\end{eqnarray}
which
represents
the mean electromagnetic force (emf) due
to the correlated fluctuations of the velocity and 
the magnetic field.
The elaboration of mean-field dynamo models  by           
Steenbeck, Krause and R\"adler \cite{KRRA} in the sixties
was a breakthrough in dynamo theory.
They showed that the mean electromotive                              
force in a non-mirrorsymmetric turbulence typically is of the form
\begin{eqnarray}
{\bf{\cal{E}}}=\alpha \overline{\bf{B}} 
- \beta \nabla                                                           
\times \overline{\bf{B}} \; ,\label{10}                                      
\end{eqnarray}
with a parameter function $\alpha$ that is 
non-zero for helical turbulence            
and a parameter $\beta$  that describes the enhancement                
of the electrical resistivity due to turbulence. The effect  
that helical fluid motion can induce
an electromotive force 
that is {\it parallel} to the magnetic field             
is now commonly known as the $\alpha$-effect.                      
Dynamo models based on the  $\alpha$-effect have played an
enormous role in the
study of solar and galactic magnetic fields, and we will
restrict all following considerations to a model of 
this kind.

Specifically, we will be concerned with the most simple
form of mean-field dynamos in spherical geometry assuming 
a spherical symmetry of the helical turbulence 
parameter, i.e.\ $\alpha{(\bf r})=\alpha(r)$, $r\in[0,1]$. Although
this is physically not realistic (for the geodynamo, 
one would find a pronounced North-South asymmetry 
of $\alpha$), the model is both simple enough for 
(quasi-)\,analytical treatments and complex enough
for allowing non-trivial back-reaction effects.  

In spherical geometry, we can decompose ${\bf{B}}$ into a
poloidal and a toroidal field component according to
${\bf{B}}=-\nabla \times ({\bf{r}} \times
\nabla {\cal S})-{\bf{r}} \times
\nabla {\cal T} $.
The defining scalar functions ${\cal S}$ and ${\cal T}$ are
then expanded
in spherical harmonics of degree $l$ and order $m$
with expansion coefficients
$s_{l,m}(r,t)$ and $t_{l,m}(r,t)$.
For the spherically symmetric 
case $\alpha{(\bf r})=\alpha(r)$ considered here,
the induction equation
decouples for each $l$ and $m$ into the pair
of equations
\begin{align}
\frac{\partial s_l}{\partial t}&=
\frac{1}{r}\frac{d^2}{d r^2}(r s_l)-\frac{l(l+1)}{r^2} s_l
+\alpha(r,t) t_l \; , \label{12}\\
\frac{\partial t_l}{\partial t}&=
\frac{1}{r}\frac{d}{dr}\left[ \frac{d}{dr}(r t_l)-\alpha(r,t)
\frac{d}{dr}(r s_l) \right]-\frac{l(l+1)}{r^2}
[t_l-\alpha(r,t)
s_l] \; , \label{13}
\end{align}
where the length is measured in units of the outer radius $R$,
the time in units of $\mu_0 \sigma R^2$, and $\alpha$ in units of 
$(\mu_0 \sigma R)^{-1}$.
The boundary conditions are given by
\begin{equation}
\label{c0}
\partial s_l/\partial r |_{r=1}+{(l+1)} s_l(1)=t_l(1)=0\;.
\end{equation}
The absence of the order $m$ in Eqs.\ \eqref{12}, \eqref{13}
follows from the spherical symmetry of $\alpha$. It implies,
in particular,
a complete degeneracy of axial and equatorial
dipole modes. It is clear that for any more realistic model
(e.g.\ with inclusion of the North-South asymmetry of $\alpha$)
this degeneration would be lifted.

\section{Spectral theory of a spherically symmetric $\alpha^2$ dynamo}

In this section, we focus on the spectral theory
of spherically symmetric $\alpha^2$ dynamos. Assuming
$\alpha{(\bf r})=\alpha(r)$ as time-independent, 
and substituting $y_1=r s_l$ and $y_2=r t_l$, we
rewrite the time evolution equation system \eqref{12}, \eqref{13}, 
\eqref{c0} 
in the previous section as an eigenvalue problem 
for a pair of coupled linear ordinary differential~equations, 
indexed by the degree $l\in\N=\{1,2,\dots\}$ of the spherical harmonics, 
\begin{align}
\label{diffsyst}
 \matrix{cc}{ \partial_r^2 - \displaystyle{\frac{l(l+1)}{r^2}} & \alpha(r) \\ 
              - \partial_r \alpha(r) \partial_r + \alpha(r)\displaystyle{\frac{l(l+1)}{r^2}} & \partial_r^2 - \displaystyle{\frac{l(l+1)}{r^2}}}
 \binom{y_1}{y_2} = \lambda \binom{y_1}{y_2}\: , \quad r\in (0,1] \; ,
\end{align}
in the product space $L_2(0,1) \oplus L_2(0,1)$ subject to the boundary condition 
\begin{align}
\label{boundcond}
\binom{(\partial_r + l ) y_1}{y_2}(1) = 0 \; .
\end{align} 
In the following, for brevity, we call the eigenvalue 
problem \eqref{diffsyst}, \eqref{boundcond} dynamo problem. 

In view of the time separation ansatz 
${\bf{B}}({\bf{r}},t)=\exp({\lambda t}) \hat{\bf{B}}({\bf{r}})$
of the corresponding magnetic field modes, 
one naturally distinguishes between decaying or 
subcritical modes \linebreak ($\Re \lambda<0$) and amplifying 
or supercritical modes ($\Re \lambda>0$), as well as between
oscillatory modes ($\Im \lambda\neq 0$) and non-oscillatory 
modes ($\Im\lambda=0$). The physically relevant 
self-sustaining dynamo configurations are
mainly defined by a few supercritical modes, whereas possible
polarity reversals of the magnetic fields are closely 
related to the existence of oscillatory modes close to criticality 
($\Re \lambda \approx 0$, $\Im \la \ne 0$) 
(see~\cite{PRL}, \cite{EPSL}).

The dynamo problem \eqref{diffsyst}, \eqref{boundcond} is not symmetric, and hence not selfadjoint, because the differential expressions 
in the off-diagonal corners of \eqref{diffsyst} are not formally adjoint to each other, they even do not have the same orders.
Therefore reliable information on the true eigenvalues of \eqref{diffsyst}, \eqref{boundcond} is very hard to obtain since
it is well-known that numerical eigenvalue approximations for non-selfadjoint problems are prone to two undesirable effects  (see e.g.\ \cite{Boegli}).
The first such effect is spectral pollution, i.e.\ eigenvalue approximations may converge to a limiting point that is not a true eigenvalue, 
a so-called spurious eigenvalue. 
The second undesirable effect is failure of spectral inclusion, i.e.\ a true eigenvalue may not be approximated.
Both effects are particularly unwanted here: a spurious eigenvalue in the half-plane $\Re \lambda>0$ could wrongly predict the existence of
supercritical modes and hence of a dynamo effect, while failure of spectral inclusion for a true eigenvalue in the half-plane $\Re \lambda>0$ 
might miss to predict the existence of a dynamo effect.

In view of this, there is an urgent need for guaranteed analytical information on the spectrum of the dynamo problem \eqref{diffsyst}, 
\eqref{boundcond} in terms of the helical turbulence function $\alpha$. A valuable tool to obtain reliable information on the location of 
eigenvalues of non-selfadjoint problems is perturbation theory for linear operators. In the simplest case, assume that $T$ is a selfadjoint (or normal) unbounded operator in  a Hilbert space
$H$ with domain $D(T)$ and $S$ is a bounded operator in $H$. Then we know that the spectrum of $T$ is real, $\sigma(T) \subset \R$, and the 
resolvent norm of $T$ satisfies 
\begin{equation}
\label{c1}
\|(T-\lambda)^{-1}\| = \frac 1 {{\rm dist} (\lambda, \sigma(T))}\: , \quad \lambda\notin \sigma(T) \;,
\end{equation}
(see \cite[Thm.\ V.3.2]{KATO}). Hence a Neumann series argument (see \cite[Ex.\ I.4.5]{KATO}) shows that for
\begin{equation}
\label{c2}
T+S-\lambda = (I_H + S(T-\lambda)^{-1})(T-\lambda)\; , \quad \lambda\notin \sigma(T) \;,
\end{equation}
the left hand side is bijective, and hence $\lambda\notin \sigma(T+S)$, provided that $\|S(T-\lambda)^{-1}\|<1$. The latter is satisfied if $\|S\|\|(T-\lambda)^{-1}\| = \|S\| / {\rm dist} (\lambda, \sigma(T))<1$ and so we obtain that
the spectrum of the perturbed operator $T+S$ satisfies the inclusion
\begin{equation}
\label{c3}
\sigma(T+S) \subset \sigma(T) + \|S\| = \{ \lambda \in \C : {\rm dist}\,(\lambda,\sigma(T)) \le \|S\|\} \;.
\end{equation}
If, however, $T$ is no longer selfadjoint (but closed, i.e.\ $T$ has closed graph), 
then the resolvent norm can no longer be controlled in terms of the distance to the spectrum, but only in terms of the distance to the numerical range 
$W(T) := \{ (Tx,x): x\in D(T), \|x\|=1\}$ of~$T$,
\begin{equation}
\label{c4}
\|(T-\lambda)^{-1}\| \le \frac 1 {{\rm dist} (\lambda, W(T))}\; , \quad \lambda\notin \overline{W(T)} \;.
\end{equation}
Note that this effect already appears in finite dimensions, the simplest example being a linear operator given by the Jordan block matrix 
\[ 
A=\begin{pmatrix} 0&1\\0&0 \end{pmatrix}, \quad 
(A-\lambda)^{-1} = \begin{pmatrix} -\lambda^{-1} & -\lambda^{-2}\\0& -\lambda^{-1}\end{pmatrix}, \ \ \lambda \notin  \sigma(A)=\{0\}\;,
\]
for which $W(A) = K_{1/2}(0)=\{z\in\C: |z|\le 1/2\}$. Clearly, the resolvent norm of $A$ does not behave like  $1 / {\rm dist} (\lambda, \sigma(A))= |\lambda|^{-1}$ as $\lambda \to 0$.

The consequence of the different resolvent norm bounds in \eqref{c1} and \eqref{c4} is that the spectrum of the perturbed operator $T+S$ now satisfies the inclusion
\begin{equation}
\label{c5}
\sigma(T+S) \subset W(T) + \|S\| = \{ \lambda \in \C : {\rm dist}\,(\lambda,W(T)) \le \|S\|\}\;.
\end{equation}
Thus the relation to the spectrum $\sigma(T)$ of the unperturbed operator $T$ is lost if $T$ is not selfadjoint!

Both spectral enclosures \eqref{c3}, \eqref{c5} may be generalized to perturbations $S$ defined on some domain $D(S) \subset H$ that are no longer bounded, 
but only relatively bounded with respect to $T$ or $T$-bounded \cite{CT16}. This means that  $D(T) \subset D(S)$  and there exist constants $a$, $b\ge 0$ with
\begin{equation}
\label{c6}
 \|Sx\| \le a \|x\| + b \|Tx\|\; , \quad x\in D(T)\;; 
\end{equation}
the infimum $\delta_T$ of all $b\ge 0$ such that \eqref{c6} holds for some $a\ge 0$ is called the $T$-bound of $S$ (see \cite[Sect.\ IV.1]{KATO}). 
Note that if $S$ is bounded, it is $T$-bounded with $T$-bound 0 since we can choose $a=\|S\|$ and $b=0$ in \eqref{c6} and that otherwise \eqref{c6} need not hold with $b=\delta_T$.

For the dynamo problem \eqref{diffsyst}, \eqref{boundcond} these perturbation results can be applied in different ways, 
depending on the choice of the unperturbed operator, to obtain rigorous and analytic enclosures of the spectrum via  \eqref{c3} or \eqref{c5}, respectively.
To this end, we first have to cast \eqref{diffsyst}, \eqref{boundcond} into a suitable operator framework.

In the Hilbert space $L_2(0,1)$ we consider the Bessel and Bessel type differential expressions, respectively, 
\begin{align} 
\label{taus}
\tau_l := -\partial_r^2 + \displaystyle{\frac{l(l+1)}{r^2}}\;, \quad 
\tau_{l,\alpha}:= - \partial_r \alpha(r) \partial_r + \alpha(r)\displaystyle{\frac{l(l+1)}{r^2}}\;,
\quad \ \partial_r:=\dfrac{{\rm d}}{{\rm d}r}\;,
\end{align}
occurring in \eqref{diffsyst} where $l\in\N=\{1,2,\dots\}$ is fixed and $\al\!:[0,1]\to\R$ is assumed to be a continuously differentiable real-valued function, $\alpha\in C^1([0,1],\R)$. 
We introduce the two Bessel differential operators $A_{l,l}$, $A_{l,\infty}$ and the Bessel type differential operator $A_{l,\alpha,\infty}$ in $L_2(0,1)$ 
by
\begin{alignat*}{3}
A_{l,l} x &= \tau_l x\;,  \quad &&   x'(1) + l x(1)=0\;, \qquad \mbox{and} \qquad 
&& A_{l,\infty} x= \tau_l x\;,   \quad x(1)=0\;, \\
A_{l,\alpha,l} x&= \tau_{l,\alpha} x\;, \quad &&  x'(1) + l x(1)=0\;,
\end{alignat*}
on respective domains $D(A_{l,l})$, $D(A_{l,\infty})$ and $D(A_{l,\alpha,\infty})$ (see \cite[Sect.~2]{UWE2} for details).
Then the dynamo problem \eqref{diffsyst}, \eqref{boundcond} can be written as an eigenvalue problem 
\[
   ({\bf A}_l - \lambda ) y = 0\;, \quad y \in D({\bf A}_l)\;,
\]
for the operator matrix ${\bf A}_l$ in the product Hilbert space $L_2(0,1) \oplus L_2(0,1)$ given by
\begin{equation}
\label{bomA_l}
 {\bf A}_l := \begin{pmatrix} -A_{l,l} & \alpha \\ \hspace{4.5mm} A_{l,\alpha,l} & - A_{l,\infty}\end{pmatrix}, 
 \quad 
 D({\bf A}_l) := D(A_{l,l}) \oplus D(A_{l,\infty})\;.
\end{equation}
In general, the spectrum of a differential operator matrix need not be discrete even if this is true for all its entries, e.g.\ 
if the product of the orders on the diagonal and the product of the orders in the off-diagonal corners are the same (see e.g.\ \cite[Sect.\ 2.4]{T}, \cite[Sect.\ 3.1]{IST16}). 
Here this does not happen, the spectrum of the operator matrix ${\bf A}_l$  remains discrete, i.e.\ it consists only of countably many eigenvalues of finite algebraic multiplicity 
without finite accumulation point.

The first decomposition of the operator matrix ${\bf A}_l$ one might think of is to split off the right upper entry,
\begin{equation}
\label{c7}
{\bf A}_l =  \begin{pmatrix} -A_{l,l} & 0 \\ \hspace{4.5mm} A_{l,\alpha,l} & - A_{l,\infty} \end{pmatrix} + 
\begin{pmatrix} 0 & \al \\ 0 & 0 \end{pmatrix}
=: T_1 + S_1\;.
\end{equation}  
The advantage of this splitting is that the operator $S_1$, whose only non-zero entry is the multiplication operator by the function $\alpha$, is bounded and that 
the spectrum of the unperturbed operator $T_1$ is known, $\sigma(T_1) = \sigma(-A_{l,l}) \cup \sigma(-A_{l,\infty}) \subset (-\infty,0)$, being the union of the eigenvalues of 
the diagonal entries, which are the zeros of certain Bessel functions of fractional order.
The disadvantage is that the unperturbed operator $T_1$ is \emph{not} selfadjoint (nor normal) and so we \emph{cannot} conclude that the spectrum of ${\bf A}_l$ lies in a 
$\|S_1\|$-neighbourhood of the eigenvalues of $-A_{l,l}$ and $-A_{l,\infty}$, and only the spectral inclusion \eqref{c5} would apply.

However, using the operator matrix structure of $T_1$ and $S_1$, we can obtain a tighter spectral enclosure. 
Indeed, rather than estimating the norm product $\|S_1\| \|(T_1-\lambda)^{-1}\|$, we
compute the product $S_1(T_1-\lambda)^{-1}$ (comp.\ \eqref{c2}) and use that it only has two non-zero entries to obtain
\begin{align*}
I+S_1(T_1\!-\!\lambda)^{-1}
=\begin{pmatrix}I-\alpha (A_{l,\infty}\!+\!\lambda)^{-1}A_{l,\alpha,l}(A_{l,l}\!+\!\lambda)^{-1}&-\alpha(A_{l,\infty}\!+\!\lambda)^{-1}\\0&I\end{pmatrix}
\end{align*}
for $\lambda \notin \sigma(-A_{l,l}) \cup \sigma(-A_{l,\infty})$. Using the relation
\begin{equation}
\label{c8}
  A_{l,\alpha,l} x = \alpha A_{l,l} x - \alpha' \partial_r x\;, \quad x\in D(A_{l,l})=D(A_{l,\alpha,l})\;,
\end{equation}
together with the resolvent estimate \eqref{c1} for the selfadjoint operators $A_{l,l}$, $A_{l,\infty}$ and the estimate $\|\partial_r A_{l,l}^{-1/2}\| \le 1$, one can derive 
the following result on the non-existence of eigenvalues of the dynamo operator ${\bf A}_l$, and hence of the dynamo problem \eqref{diffsyst}, \eqref{boundcond} 
(see \cite[Thm.~4.6 and Cor.~4.8]{UWE2}). 

Here, for a continuous function $f\in C([0,1],\C)$ we denote by $\|f\|_{\infty}:=\max_{r\in[0,1]} |f(r)|$ its maximum norm.

\smallskip

\hspace{-4.5mm}{\bfseries{Anti-dynamo theorem.}}
\emph{The dynamo operator ${\bf A}_l$ associated with the $\alpha^2$ dynamo problem {\rm \eqref{diffsyst}, \eqref{boundcond}} possesses no eigenvalues with real part $>0$ if 
$\alpha\in C^1([0,1],\R)$ and
\begin{equation}
 \label{anti}
\|\alpha\|_{\infty}^2 + \dfrac{\|\al\|_{\infty}^2\|\alpha'\|_{\infty}^2}{j_{l-\frac 12,1}}< j_{l+\frac 12,1}^2
\end{equation}
where $j_{l-\frac 12,1}$ and $j_{l+\frac 12,1}$ 
are the smallest non-zero zeros of the two Bessel functions $J_{l-\frac 12}$ and $J_{l+\frac 12}$, respectively.
The full dynamo operator ${\bf A} = \bigoplus_{l=1}^\infty {\bf A}_l$ possesses no eigenvalues with real part $>0$ if \eqref{anti} holds for $l=1$.
}

\smallskip

Note that the last claim follows from the interlacing property $j_{l-\frac 12,1} < j_{l+\frac 12,1}$ of the Bessel zeros for $l\in \N$ (see \cite[9.5.2 and 10.1]{as}) which implies that 
$j_{\frac 12,1} \le j_{l-\frac 12,1}$ and $j_{\frac 32,1} \le j_{l+\frac 12,1}$ for $l\in \N$.

\smallskip

{\hspace{-4.5mm}{\bfseries{Example 1.}} 
For the simplest profile $\alpha_{kin}(r)=C$, $r\in[0,1]$, the eigenvalue problem~\eqref{diffsyst}, \eqref{boundcond}
is analytically solvable (see e.g.\ \cite{KRRA}) to give, for $l\!=\!1$, a critical value of $C\!=\!j_{\frac 32,1}$ $(\sim\!4.493409)$.
In this particular case, our general anti-dynamo criterion \eqref{anti} specializes~to
\[
\|\alpha\|_{\infty}^2=C^2<j^2_{\frac 32,1} \ (\sim 4.493409^2)\;,
\]
which perfectly matches the analytically known critical value for $C$ and shows that \eqref{anti} is sharp in this case.

\smallskip

\hspace{-4.5mm}{\bfseries{Example 2.}}
For the kinematic profile $\alpha_{kin}(r)=1.916 \cdot C  \cdot (1-6 \; r^2+5 \; r^4)$, $r\in[0,1]$ (which
will serve as a paradigmatic profile for the reversal studies in the next section), 
it is not difficult to check that
\[
  \|\alpha\|_{\infty} = \alpha(0)= 1.916 \cdot C\;, \quad \|\alpha'\|_{\infty} = \alpha'(1) = 8 \cdot 1.916 \cdot C\;.
\]
The Bessel zeros in \eqref{anti} for $l=1$ satisfy $j_{\frac 12,1} = \pi \le 3.142$ and $j_{\frac 32,1} \le 4.494$, respectively, and so \eqref{anti} for $l=1$ holds if 
\[
 1.916^2 \cdot C^2 \Big( 1 + \frac 8{3.142} \Big) < 4.494^2\,.
\]
Hence it is analytically guaranteed that for $\alpha_{kin}$ with $C < 1.725$ the dynamo problem {\rm \eqref{diffsyst}, \eqref{boundcond}} possesses no eigenvalues with real part $>0$ for any $l\in\N$.

\smallskip

While the above decomposition \eqref{c7} shows, more generally, that all eigenvalues lie in a half-plane $\Re \lambda \le a_l$ with some constant $a_l \in \R$ 
and provide good control of the real parts of all eigenvalues, it does not provide a good estimate for the imaginary parts. In fact, it only yields an estimate of the form 
$|\Im \lambda | \le h_l (\Re \lambda)$ for $\Re \lambda \in (-\infty,a_l]$  with a continuous strictly decreasing function $h_l\!:(\!-\infty,a_l] \to [0,\infty)$ with
$h_l(a_l)=0$, but $\lim_{t\to-\infty} h_l (t) = \infty$ and $\lim_{t\nearrow a_l} h'_l (t)=-\infty$. So in particular, it does not exclude that the 
imaginary parts of the eigenvalues tend to $\pm \infty$ when their real parts tend to $-\infty$.

In order to obtain a better estimate for the imaginary parts of the eigenvalues, and possibly information on the number of non-real eigenvalues, a more subtle decomposition
of the dynamo operator ${\bf A}_l$ is needed, preceded by a quasi-similarity transformation with the (unbounded) operator matrix 
\[
{\mathcal W} := \begin{pmatrix} A_{l,l}^{1/2}& 0 \\ 0 & I\end{pmatrix},\quad 
D(\mathcal W) := D(A_{l,l}^{1/2}) \oplus L_2(0,1)\;.
\]
Although the transformation is not bounded, the spectra of ${\bf A}_l$ and of the transformed operator ${\mathcal W}^{-1} {\bf A}_l {\mathcal W}$ 
coincide since they consist only of eigenvalues.  Invoking the relation \eqref{c8} again, we see that 
\begin{equation}
\label{c9}
{\mathcal W}^{-1} {\bf A}_l {\mathcal W} 
= \begin{pmatrix} -A_{l,l} \!&\! A_{l,l}^{1/2} \alpha  \\  \alpha A_{l,l}^{1/2} \!&\! - A_{l,\infty} \end{pmatrix}
+ \begin{pmatrix} 0 & 0 \\[-1mm] -\alpha' \partial_r A_{l,l}^{-1/2} \!&\! 0 \end{pmatrix} =: T_2 + S_2\;.
\end{equation}
The advantage of this decomposition is that here the unperturbed operator $T_2$ is selfadjoint and at the same time the perturbation $S_2$ is bounded (since $\|\partial_r A_{l,l}^{-1/2}\| \le 1$) with
$\|S_2\| \le \|\alpha'\|_{\infty}$; by \eqref{c3} this immediately implies that the imaginary parts of all eigenvalues of the dynamo operator ${\bf A}_l$ are uniformly bounded by $ \|\alpha'\|$, 
\begin{equation}
\label{c10}
|\Im \lambda | \le  \|\alpha'\|_{\infty} \quad \mbox{for all } \lambda \in \sigma({\bf A}_l)\;;
\end{equation}
note that \eqref{c10} readily yields that for constant $\alpha$ all eigenvalues of the dynamo problem \eqref{diffsyst}, \eqref{boundcond} are real.

The disadvantage of the decomposition \eqref{c9} is that we no longer have explicit information on the spectrum of the unperturbed operator $T_2$. 
Nevertheless, $T_2$ is a \emph{diagonally dominant} operator matrix (see \cite[Def.\ 2.2.1]{T}) and we can decompose the operator $T_2$ into its dominating diagonal and the off-diagonal part, 
\[
 T_2 = \begin{pmatrix} -A_{l,l} \!&\! 0 \\  0 \!&\! - A_{l,\infty} \end{pmatrix}
       + \begin{pmatrix} 0 \!&\! A_{l,l}^{1/2} \alpha  \\  \alpha A_{l,l}^{1/2} \!&\! 0 \end{pmatrix} =: D_2 + O_2\;.
\]
Here both the diagonal part $D_2$ and the off-diagonal part $O_2$ are selfadjoint and the spectrum
$\sigma(D_2) = \sigma(-A_{l,l}) \cup \sigma(-A_{l,\infty})$ is given by  the two interlacing sequences of non-zero Bessel zeros $-\lambda_k(l,l)=-j_{l-\frac 12,k}^2$ of 
$\la \mapsto J_{l-\frac 12}(\sqrt{\la})$ and  $-\lambda_k(l,\infty) = -j_{l+\frac 12,k}^2$ of $\la \mapsto J_{l+\frac 12}(\sqrt{\la})$, respectively, for $k\in\N$,
\begin{equation}
\label{c11}
 \sigma(D_2) = \{ -\lambda_k(l,l) \}_{k=1}^\infty \cup \{ -\lambda_k(l,\infty) \}_{k=1}^\infty = \{ - j_{l-\frac 12,k}^2 \}_{k=1}^\infty \cup \{ - j_{l+\frac 12,k}^2 \}_{k=1}^\infty\;,
\end{equation}
and all eigenvalues of $D_2$ are simple. Further, it can be shown that the off-diagonal part $O_2$ is $D_2$-bounded and constants $a_{\alpha}$, $b_{\alpha}\ge 0$ in the respective inequality \eqref{c6} may be found explicitly 
in terms of the helical turbulence function $\alpha$.
Applying a more advanced perturbation result (see \cite[Thm.\ 2.1]{CT16}), one can establish estimates for the eigenvalues of $T_2$ which guarantee that the eigenvalues $\{\mu_k(t)\}_{k=1}^\infty$ of the selfadjoint operator
$D_2+ t O_2$ for $t\in [0,1]$ vary in intervals around the eigenvalues of $D_2$ that remain disjoint for all $t\in [0,1]$. In particular, this guarantees that the eigenvalues $\{\mu_k(1)\}_{k=1}^\infty$
of $T_2=D_2+ O_2$ are still all simple. Moreover, after some tedious and clever analysis, one can estimate their differences to the eigenvalues of $D_2$, i.e.\ 
the Bessel zeros in \eqref{c11} in terms of the relative boundedness constants $a_{\alpha}$, $b_{\alpha}\ge 0$.

These involved eigenvalue estimates may be combined with the asymptotics of the Bessel zeros forming the spectrum of $D_2$, and with the following operator theoretic property. 
Since the helical turbulence function $\alpha$ is real-valued, the eigenvalues of the dynamo operator ${\bf A}_l$, and hence of the operator $T_2 + S_2$ in \eqref{c9} are symmetric to the real axis.
If it is guaranteed that the eigenvalues of the selfadjoint operator $T_2$ are all simple, then after the perturbation $S_2$ eigenvalues may only become complex (and hence form a complex conjugate pair) 
if two eigenvalues coalesce (comp.\ \cite{LT06}). The eigenvalue estimates for $T_2$ together with the fact that $S_2$ is bounded finally yield the following result (see \cite[Thm.\ 3.13]{Kaeser17}).

\smallskip

\hspace{-4.5mm}{\bfseries{Theorem.}}
\emph{If $\alpha \in C^2([0,1],\R)$  and $\|\alpha\|_{\infty} < \pi/  2^{5/4}$, then the dynamo problem {\rm \eqref{diffsyst}, \eqref{boundcond}} has 
at most a finite number of non-real $($complex conjugate pairs of\,$)$ eigenvalues.}

\smallskip

We mention that the norm bound on $\alpha$ guarantees the simplicity of the eigenvalues of $T_2$, so we expect it is neither sharp nor even necessary;
e.g.\ for the kinematic profile $\alpha_{kin}(r)=1.916 \cdot C  \cdot (1-6 \; r^2+5 \; r^4)$, $r\in[0,1]$, the bound is satisfied if $C <0.689$.
Note also that the derivative $\alpha'$ 
will play a role if one is interested in estimating the number of non-real eigenvalue pairs (it is zero if $\alpha'=0$, i.e.\ if $\alpha$ is constant).

It should be emphasized that it is \emph{not} possible to 
obtain a result on the finiteness of the number of non-real 
eigenvalues by means of numerical computations, simply 
because - even if they were guaranteed computations
e.g.\ with interval arithmetic (see \cite{BLMTW10,BBMTW14}) - they 
can only cover bounded regions of the complex plane in finite time.

\section{A simplified reversal model for the geodynamo }

In the previous section we have seen that spherically symmetric
$\alpha^2$ dynamos may have complex eigenvalues as long as the
radial profile $\alpha(r)$ is not too simple 
(e.g.\ for constant $\alpha$ there are only real eigenvalues). 
Now we assume that the profile $\alpha(r)$ in the 
kinematic regime, $\alpha_{kin}(r)$,
leads to dynamo action which might be either of 
non-oscillatory or oscillatory type.
After self-excitation, i.e.\ exponential magnetic field growth, 
has set in, saturation is ensured by a reduction, or 
``quenching'', of $\alpha$. This effect can be interpreted 
as a realization of  Lenz's rule, according to which 
the magnetic field acts against the source of its own
generation.

Actually, this quenching is a non-trivial
and strongly debated
mechanism. In order to remain within the framework of spherically 
symmetric $\alpha^2$ dynamos, we take resort to the 
simplification  that the quenching of the $\alpha$ profile
is affected by the angularly averaged
magnetic field energy which
can be expressed in terms of $s(r,t)$ and $t(r,t)$. Of course, a realistic 
quenching would introduce terms breaking
the spherical symmetry  of $\alpha$.

In addition to the quenching effect, we
will later consider the $\alpha$ profile to
be further affected by some noise terms which are
supposed to be constant within a correlation
time $\tau_{corr}$.
Physically, this noise could
be understood as a consequence of changing  
boundary conditions for the
core flow, but also as a  substitute
for the omitted
influence of higher multipole modes on the dominant 
dipole mode.

Putting these two effects together, the $\alpha$ profile 
takes on the time dependent form
\begin{eqnarray}
\label{eqquench}
\alpha(r,t)=C \; \frac{\alpha_{kin}(r)}{1+
E_{mag}(r,t)/E^0_{mag}}+\Xi(r,t) \; ,
\end{eqnarray}
where
$E_{mag}$ is the magnetic energy, averaged over the
angles,
\begin{eqnarray}
E_{mag}(r,t)=\frac{2 s_1^2(r,1)}{r^2}+
\frac{1}{r^2}\left( \frac{\partial (r s_1(r,t))}
{\partial r} \right)^2
+t_1^2(r,t) \; .
\end{eqnarray}
In the numerical scheme, the noise term $\Xi(r,t)$
will be treated in form of a Taylor expansion,
\begin{eqnarray}
\Xi(r,t)=\xi_1(t) +\xi_2(t)
\; r^2 +\xi_3(t) \; r^3+\xi_4(t) \; r^4 \; ,
\end{eqnarray}
with  the noise correlation given by
$\langle \xi_i(t) \xi_j(t+t_1)
\rangle = D^2 (1-|t_1|/\tau_{corr})
\Theta(1-|t_1|/\tau_{corr}) \delta_{ij}$.

In summary, our model is governed by
four parameters: the
magnetic Reynolds number $C$, the noise
amplitude $D$, the noise
correlation time $\tau_{corr}$, and
the mean magnetic energy $E^0_{mag}$
in the saturated regime.

The system of equations \eqref{12}, \eqref{13} and \eqref{eqquench} 
is time-stepped
using an Adams-Bashforth method.
For the following examples,
the correlation time $\tau_{corr}$ has been set to 0.02,
and $E^0_{mag}$ has been
chosen to be 100. The details of these choices are not
very relevant. Roughly speaking, a shorter correlation
time $\tau_{corr}$ would require a stronger
noise amplitude $D$ in order to yield the same effect.

In the remainder of this section, we analyze in detail 
a dynamo which starts from the kinematic profile 
$\alpha_{kin}(r)=1.916 \cdot C  \cdot (1-6 \; r^2+5 \; r^4)$, $r\in[0,1]$, 
(the prefactor 1.916 merely serves for normalizing the 
profile's intensity to 1).
This profile $\alpha_{kin}$, 
with its sign change along the radius (see the 
full line in Fig.\ \ref{Fig:relax}\,(b)), 
is a good candidate to exhibit complex eigenvalues
and, therefore, an oscillatory dynamo behaviour.
Indeed, the dynamo (upper panel of Fig.\ \ref{Fig:relax}\,(a))
arising at  $C=6.8$, which is 
only slightly larger than the critical value of 6.78, 
turns out to be of oscillatory character (we note in passing that
the critical value of 6.78 is significantly larger than the 
safe limit of 1.725 for the non-existence of 
eigenvalues with positive real part, as derived in the previous 
section, since the latter holds for any $\alpha \in C^1((0,1),\R)$). While the oscillation 
at this value of $C$ is nearly harmonic, a strongly anharmonic 
oscillation shows up already at the slightly larger value 
$C=7.237$ (the similarity 
of this anharmonic oscillation with the relaxation oscillation 
of the van-der-Pol oscillator had been discussed in \cite{GAFD}).
Increasing $C$ just a little further to $C=7.24$,  this strongly 
anharmonic oscillation turns into a steady dynamo. 
For all higher values of $C$ considered here, 
the dynamo remains steady.

\begin{figure}
\includegraphics[width=13cm]{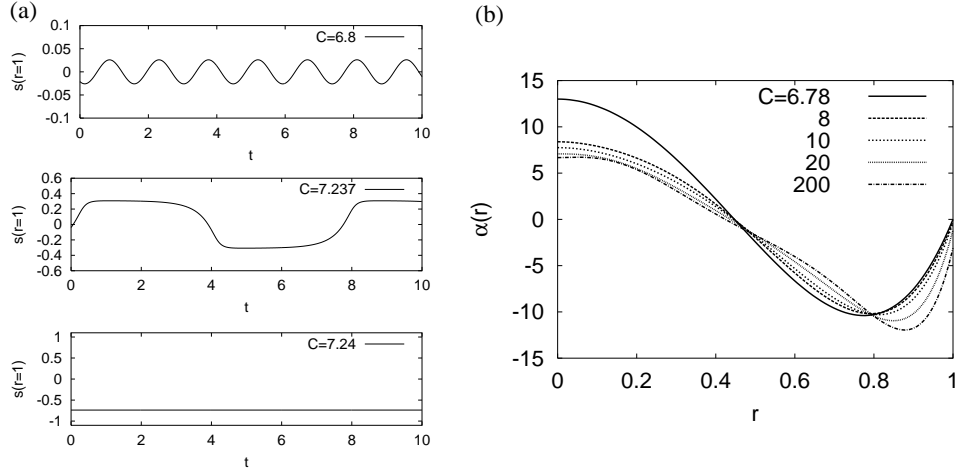}
\caption{Typical time series of 
the $\alpha^2$ dynamo model for three different values of $C$ (left). 
With increasing $C$  one observes an increasing
degree of anharmonicity which finally turns
into a steady state. Right: (nearly) kinematic and four 
saturated $\alpha$ profiles.}
\label{Fig:relax}
\end{figure}

For a few selected values of $C$, the quenching effect of the 
self-excited magnetic
field on the $\alpha$ profile is illustrated in 
Fig.\ \ref{Fig:relax}\,(b).
Evidently, all initial kinematic profiles $\alpha_{kin}$
with increasing $C$
are quenched into saturated profiles which are actually quite 
similar in shape and intensity. However, these comparably 
minor changes of the shape of the profile 
lead to a drastic change from oscillatory
(for $C=6.8$) to non-oscillatory behaviour (for 
$C>7.24$).

Can we understand this saturated dynamo behaviour in terms 
of spectral theory? In the previous section we learned 
that in our problem 
eigenvalues may only form complex conjugate pairs 
if two real eigenvalues have coalesced at an exceptional 
point. This is illustrated in Fig.~\ref{Fig:spectrum}
which visualizes the 
relevant part of the spectrum for four $\alpha$ 
profiles corresponding to the (nearly) unquenched profile for  
$C=6.78$ and to the quenched profiles resulting for $C=8$,~$20$ and $200$. 
Note that here the abscissa is labeled by an artificial factor~$C^*$ (different from $C$!) which scales the {\it quenched} profile
(therefore, $C^*=1$ always corresponds to the 
point of marginal, saturated dynamo action with zero growth rate, 
while lower and
higher values of $C^*$ just serve to illustrate
the spectral behaviour around this point).

These spectra are very telling: while the spectrum for $C=6.78$ 
exhibits an exceptional point below the threshold of dynamo action 
(i.e.\ for $C^*<1$), and a complex eigenvalue at $C^*=1$, 
the situation changes dramatically
for higher values of $C$. For $C=8$ the exceptional point 
has shifted into the right upper part
so that for $C^*=1$ we obtain a non-oscillatory 
dynamo (see the steady dynamo for $C=7.24$ as shown in 
Fig.\ \ref{Fig:relax}\,(a)). Remarkably, for the even
higher values $C=20$ and 200, the exceptional point 
moves back to a value close to $C^*=1$. For $C=200$, 
it even seems to 
``cling'' to the zero growth rate line, just as 
if this highly super-critical dynamo would like to
develop an exceptional point close
to its marginal point. We 
will come back to this feature further below.

\begin{figure}
\includegraphics[width=13cm]{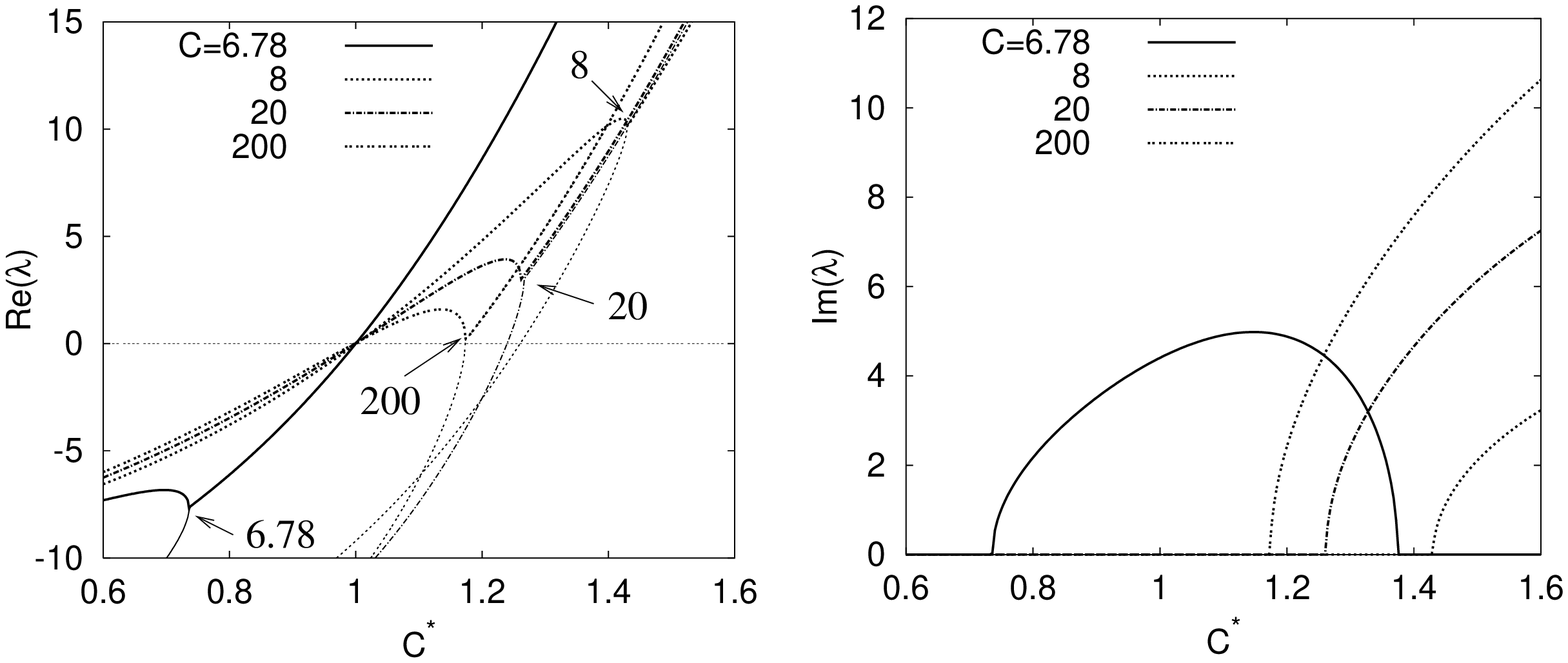}
\caption{Spectral properties of the nearly kinematic ($C=6.78$)
and of the saturated ($C=$8, 20, 200) $\alpha$ profiles which
result from the choice
$\alpha_{kin}(r)=1.916 \cdot C  \cdot (1-6 \; r^2+5 \; r^4)$, 
$r\in[0,1]$.
The scaling with the artificial factor $C^*$  helps to identify the
actual eigenvalue (at $C^*=1$) in its relative position to the
exceptional point. Note that for highly supercritical $C$ the
exceptional point moves close to the zero growth rate line.
Left: growth rate: Right: angular frequency.}
\label{Fig:spectrum}
\end{figure}

Up to this point, the time evolution was supposed to be 
completely deterministic which led, depending on the value of $C$,
to (nearly harmonic or anharmonic) oscillations or to steady dynamos. 
The spectral portraits of Fig.\ \ref{Fig:spectrum} indicate 
that for high values of $C$ the exceptional point
has a tendency to move very close to the marginal 
point, which makes it likely that even weak noise 
leads to transitions between non-oscillatory and 
oscillatory branches. 
In the following we 
relate those transitions to the reversals
of the (geo-)\,dynamo field.

This is exemplified in Fig.\ \ref{Fig:noise} 
which shows the time evolution for 
two values of $C$, combined with two values of the 
noise intensity $D$.
Evidently, for $C=20$ the noise strength $D=5$ 
only leads  to some 
fluctuations of the field while it is not sufficient 
to trigger any reversals. The latter show up, 
however, for the slightly 
higher noise level $D=6$. 
Interestingly, the stronger dynamo with $C=50$ 
already exhibits  reversals for
$D=5$ which can be attributed to the closer proximity of 
the exceptional point to the
zero growth rate line. 
For $C=50$ and $D=6$, the rate of field 
reversals becomes quite high.

\begin{figure}
\includegraphics[width=13cm]{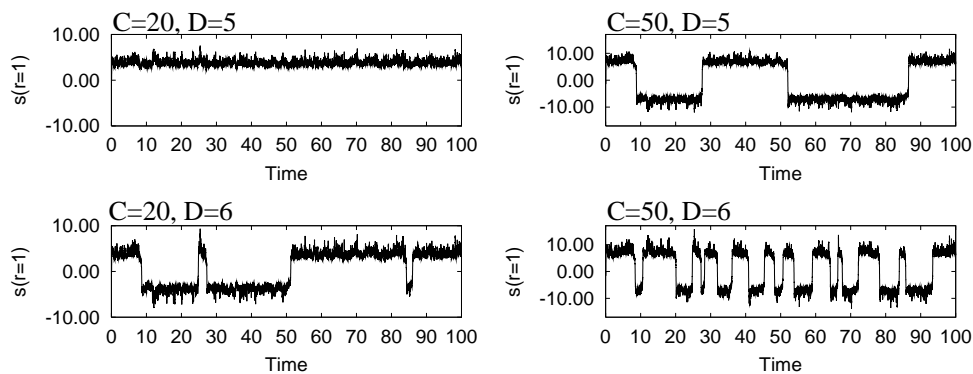}
\caption{\!Time series for various values of $C$\!, $D$ and  kinematic profile 
$\alpha_{kin}\!(r)\!=\!1.916\,C(1\!-\!6r^2\!+\!5r^4)$.}
\label{Fig:noise}
\end{figure}

Fig.\ \ref{Fig:scheme} provides a schematic 
explanation of reversals as 
noise-triggered relaxation oscillations in 
the vicinity of exceptional points.
Actually, this scheme represents a spectral theoretical 
counterpart of the dynamical systems picture 
developed in \cite{PETRELIS} 
which interprets reversals
in terms of  a saddle node bifurcation
where a pair of a stable and an unstable fixed point 
is replaced
by a Hopf bifurcation.
In our Fig.\ \ref{Fig:scheme}, the crossing point S of the 
real spectral branch with the zero growth rate line 
represents a stable fixed point for which 
any slight changes to the left or right 
would always act  in a stabilizing manner. 
For the unstable fixed point 
U the situation is different; here, due
to the negative slope of the real branch, 
any change leads to a repelling from this point
(in close analogy with the van-der-Pol 
oscillator with its typical ``negative differential resistance region'').
The role of noise is to bring the system
from the stable fixed  point S ``over the hill" (around the local 
maximum M) towards the unstable fixed point U, from
where it is forced into the oscillatory branch where
the very sign change (i.e.\ where the reversal) occurs, before the
system settles again into a stable fixed point.

\begin{figure}
\begin{center}
\includegraphics[width=10cm]{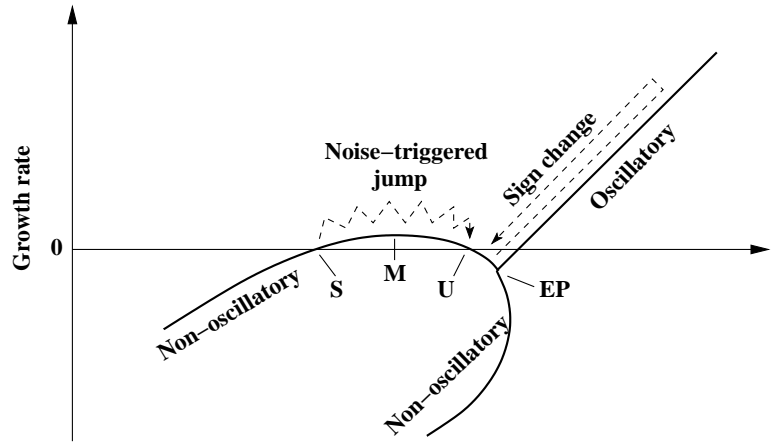}
\caption{Illustration of the main mechanism and the
various phases of a reversal in the vicinity of an
exceptional point of the spectrum of a saturated,
highly supercritical dynamo.
}
\end{center}
\label{Fig:scheme}
\end{figure}

In Fig.\  \ref{Fig:vadm} we evidence the remarkable similarity of the
reversal process resulting from our model with the 
typical reversals revealed by paleomagnetic measurements. For the parameter combination 
$C=20$, $D=6$,
Fig.\ \ref{Fig:vadm}\,(a) exemplifies five typical reversals
and their average (note that the time scale has been
adjusted to the typical geophysical values). 
For the sake of comparison, Fig.\ \ref{Fig:vadm}\,(b) 
shows the last five paleomagnetic
reversals and their average. Typical for either time series 
is the well-expressed asymmetry of the reversals, comprising  
a slow decay and a fast recovery.

\begin{figure}
\includegraphics[width=13cm]{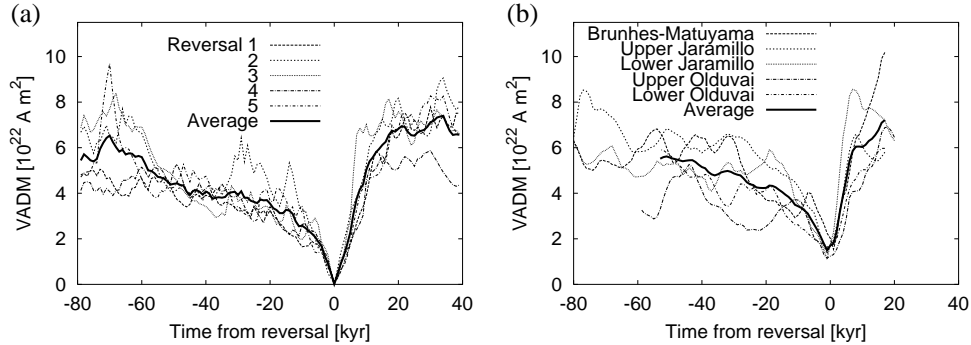}
\caption{Comparison of  
numerically simulated reversals and paleomagnetic reversal data. 
(a) Five typical reversals resulting from the dynamo model
with $\alpha_{kin}(r)=1.916 \cdot C  \cdot (1-6 \; r^2+5 \; r^4)$, $r\in[0,1]$,
for $C=20$, $D=6$, and their average.
The time scale and the scale of the virtual axial dipole moment (VADM) 
have been chosen in such a way that they become comparable to 
the geophysical data. Note that the negative VADM values after the reversal
have been mirrored to the positive side. 
(b) VADM during
the 80 kyr preceding  and the 20 kyr following a polarity transition
for five reversals from the last 2 million years 
(data extracted from \cite{VALET}),
and their average.}

\label{Fig:vadm}
\end{figure}

In \cite{INVERSE} we have pursued this comparison up to the 
point that we identified (by means of a simplex method)
a number of essential
parameters of the geodynamo by matching a synthetic reversal 
process to various paleomagnetic reversal data.  
In doing so, we have carefully avoided any over-interpretation 
of our simple model by  focusing  only on those 
parameters to be determined, and those 
functionals to be minimized, that refer to the  temporal properties of 
reversal sequences, and not to any spatial features. 
This makes us optimistic that the results will prove robust 
when  inversions of this kind will later be
repeated using more realistic dynamo models.

\section{Conclusions and prospects}

In this paper, we have related geomagnetic reversals to the
specific spectral properties of non-selfadjoint dynamo 
operators of mean-field type. For this purpose, we 
have analyzed  in detail the spectra of spherically symmetric
$\alpha^2$ dynamos and have derived strict criteria
for the occurrence of complex eigenvalues.

Then we have shown that 
the typical asymmetry of reversals
can be explained as a noise
triggered relaxation oscillation in the vicinity 
of an exceptional point of the spectrum where two real
branches coalesce to form one common complex branch. 
While from a purely kinematic viewpoint the occurrence of an 
exceptional point close to the zero growth rate line
appears as rather accidental (and hence of no 
explanatory use), things are different for the
saturated states of highly supercritical dynamos.

For a typical example with a sign change along the radius
we have shown that with increasing dynamo strength 
the modified, saturated $\alpha$ profile
develops spectra that become more and more prone 
to reversals. It is tempting to interpret this 
mechanism as a sort of self-organized criticality, 
although more work would be needed to support this
claim.

A necessary next step would be to apply our methods to 
more realistic dynamo models.  The three-dimensional
structure of $\alpha$ as it results from full 3D simulations 
of the geodynamo and which also shows the typical
sign change along the radius \cite{GIESECKE1,SCHAEFFER}
makes this a promising enterprise.

\begin{acknowledgement}
  We thank Uwe G\"unther for many years of fruitful 
   collaboration on various aspects of  
   the spectral theory of non-selfadjoint operators (F.S.) and 
   the physics of dynamos (C.T.), 
   and for bringing the two authors of this paper together.
\end{acknowledgement}

%
%

\end{document}